\documentclass[aps,prb,twocolumn,reprint,nofootinbib,superscriptaddress,showkeys]{revtex4-2}

\usepackage{color}
\usepackage{graphicx}
\usepackage{dcolumn}
\usepackage{bm}
\usepackage{commath}
\usepackage{amsmath}
\usepackage{upgreek}
\usepackage{siunitx}
\usepackage{physics}
\usepackage{natbib}
\usepackage[english]{babel}
\usepackage[latin1]{inputenc}
\usepackage[bottom=2cm,top=2cm, left=1.5cm, right=1.5cm]{geometry}
\usepackage{booktabs}
\usepackage[unicode=true, bookmarks=true, bookmarksnumbered=false, bookmarksopen=false, breaklinks=false, pdfborder={0 0 1}, backref=false, colorlinks=false]{hyperref}
\usepackage[normalem]{ulem}
\hypersetup{pdftitle={Terahertz crystal electric field transitions in a Kondo-lattice antiferromagnet}, pdfauthor={Payel Shee, Chia-Jung Yang, Ruta Kulkarni, Arumugam Thamizhavel, Manfred Fiebig, and Shovon Pal}}

\begin{document}

\title{Terahertz crystal electric field transitions in a Kondo-lattice antiferromagnet}

\author{Payel Shee}
\altaffiliation{Both authors contributed equally}
\affiliation{School of Physical Sciences, National Institute of Science Education and Research, An OCC of HBNI, Jatni, 752 050 Odisha, India}

\author{Chia-Jung Yang}
\altaffiliation{Both authors contributed equally}
\affiliation{Department of Materials, ETH Zurich, 8093 Zurich, Switzerland}

\author{Shishir Kumar Pandey}
\affiliation{Artificial Intelligence for Science Institute, 100 080 Beijing, China}

\author{Ashis Kumar Nandy}
\affiliation{School of Physical Sciences, National Institute of Science Education and Research, An OCC of HBNI, Jatni, 752 050 Odisha, India}

\author{Ruta Kulkarni}
\affiliation{Department of Condensed Matter Physics and Material Science, Tata Institute of Fundamental Research, 400 005 Mumbai, India\looseness=-1}

\author{Arumugam Thamizhavel}
\affiliation{Department of Condensed Matter Physics and Material Science, Tata Institute of Fundamental Research, 400 005 Mumbai, India\looseness=-1}

\author{Manfred Fiebig}
\affiliation{Department of Materials, ETH Zurich, 8093 Zurich, Switzerland}

\author{Shovon Pal}
\altaffiliation{shovon.pal@niser.ac.in}
\affiliation{School of Physical Sciences, National Institute of Science Education and Research, An OCC of HBNI, Jatni, 752 050 Odisha, India}

\date{\today}

\begin{abstract}
Hybridization between the localized $f$-electrons and the delocalized conduction electrons together with the crystal electric field (CEF) play a determinant role in governing the many-body ground state of a correlated-electron system. Here, we investigate the low-energy CEF states in CeAg$_{2}$Ge$_{2}$, a prototype Kondo-lattice antiferromagnet where Kondo correlation is found to exist within the antiferromagnetic phase. Using time-domain THz reflection spectroscopy, we show the first direct evidence of two low-energy CEF transitions at 0.6\,THz (2.5\,meV) and 2.1\,THz (8.7\,meV). The presence of low-frequency infrared-active phonon modes further manifests as a Fano-modified lineshape of the 2.1\,THz CEF conductivity peak. The temporal spectral weights obtained directly from the THz time traces, in addition, corroborate the corresponding CEF temperature scales of the compound.
\end{abstract}

\maketitle

\section{Introduction}
Intermetallic compounds provide a rich platform for the realization of diverse electronic ground states. This unique aspect results from the competition between the Ruderman-Kittel-Kasuya-Yosida (RKKY)~\cite{Ruderman1954,Kasuya1956,Yosida1957} and the Kondo~\cite{Kondo1964,Hewson1993,Kirchner2020} interactions. While the RKKY interaction promotes a magnetically ordered ground state, the Kondo effect leads to a disordered paramagnetic state. The competition further leads to the emergence of exotic properties at low temperatures, some of which are unconventional superconductivity, heavy-fermion behavior and its coexistence with the magnetic phase. The Ce-based intermetallic compounds with the molecular formula CeT$_2$X$_2$ ($\text{T}$ = transition metal and $\text{X}$ = semiconductor), have been widely investigated to realize the emerging phenomena such as heavy-fermion superconductivity~\cite{Steglich1979,Ren2014}, pressure-induced superconductivity~\cite{Sheikin2001,Movshovich1996}, and unconventional metamagnetic transitions~\cite{Haen1987} along with fundamental investigations of thermodynamic and transport properties~\cite{Bohm1988,Endstra1993,Singh2012}. The magnetic moments in these compounds are found on the Ce $f$-orbitals instead of the transition-metal ions. As we lower the temperature of the material, these localized $f$-states readily hybridize with the spatially extended $s$-, $p$- or $d$-states that form the conduction band~\cite{Andres1975,Stewart1984}. The hybridization along with the crystal electric field (CEF) dictate the low-temperature magnetic phase~\cite{Babkevich2015,Sundermann2018}. Here, the CEF is the electrostatic field produced by the ligands present in the crystal structure, resulting from the electrostatic interaction between the central magnetic ion and the ligand ions surrounding it. Depending on the nature of the hybridization, the ground-state properties of the material can vary from the Kondo-dominated magnetically disordered state to the RKKY-mediated magnetically ordered state~\cite{Si2001,Kouwenhoven2001}.

\begin{figure*}[t!]
	\centering
	\includegraphics[width=0.7\linewidth]{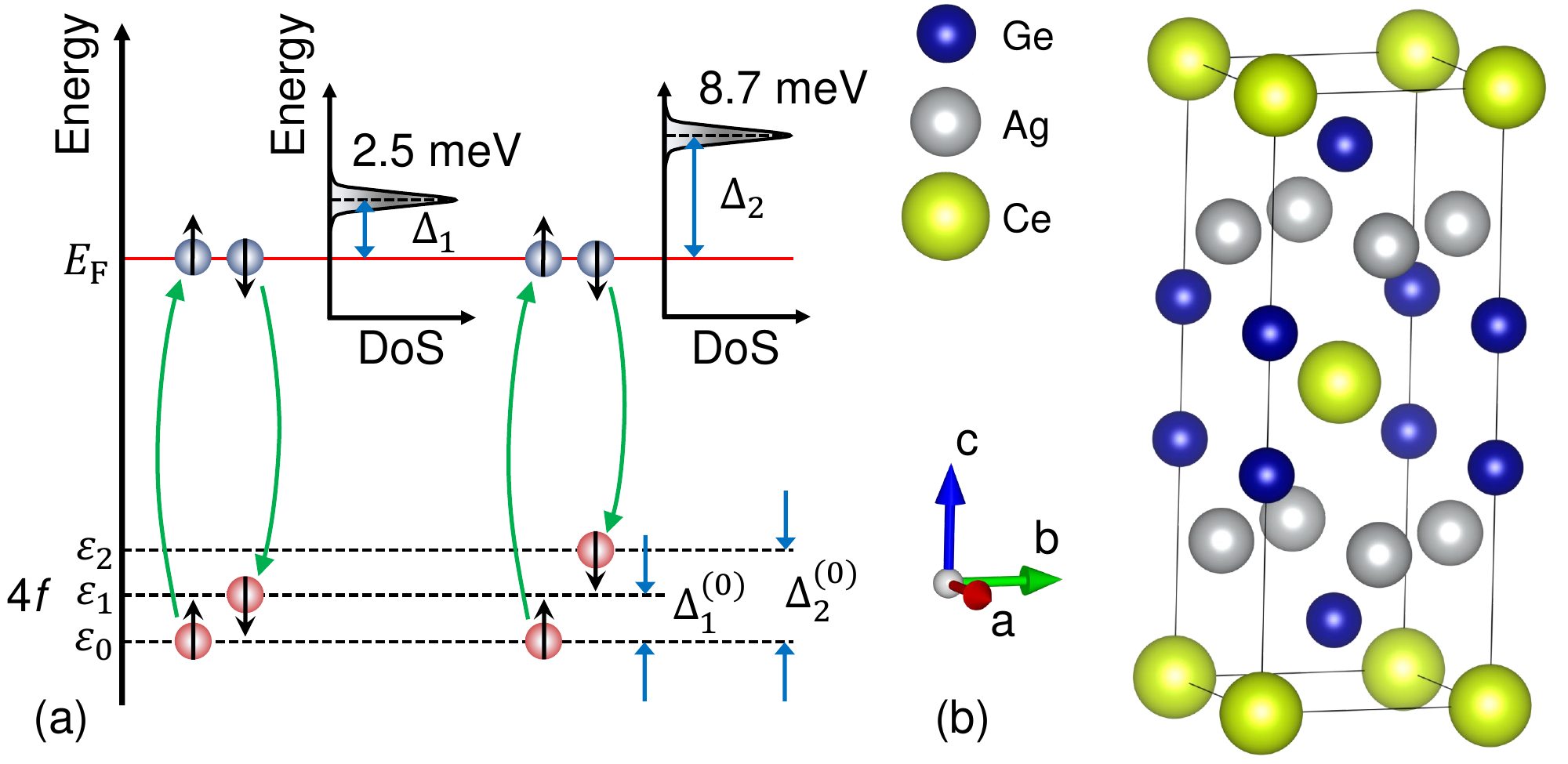}
	\caption{(a) Schematic diagram that shows the spin-flip transition of the 4$f$-electrons giving rise to the narrow CEF resonance peaks in the density of states (DoS) near the Fermi level $E_{\rm F}$. $\varepsilon_0$, $\varepsilon_1 $, and $\varepsilon_2$ are the Kramers doublets. $\Delta_{1,2}^{(0)}$ are the bare CEF excitation energies while $\Delta_{1,2}$ are the renormalized CEF energies that are measured in our experiments and correspond to the transitions from $\varepsilon_0 \to \varepsilon_{1,2}$ (CEF$_{1,2}$), respectively. (b) The body-centered tetragonal crystal structure of CeAg$_{2}$Ge$_{2}$.}
	\label{fig1}
\end{figure*}

The heavy-fermion behavior is perhaps the most attractive and widely investigated phenomenon in Ce-based intermetallics, where the hybridization results in the formation of heavy quasiparticles at low temperatures~\cite{Coleman2006}. This concomitantly enhances the density of states at the Fermi energy $E_{\rm F}$ resulting in an enlarged Fermi surface~\cite{Kummer2015,Pal2019}. In the absence of hybridization, the magnetic properties of Ce-based heavy-fermion compounds are governed by the local environment of the Ce ion. The $J = 5/2$ ground-state multiplet of the Ce-4$f$ states has six-fold degeneracy that, however, is lifted by the CEF into three Kramers doublets, in a non-cubic environment, denoted by $\varepsilon_0$, $\varepsilon_1$, $\varepsilon_2$ (see Fig.~\ref{fig1}a). Due to the hybridization of the 4$f$ states with the conduction-electron states, a Ce 4$f$ electron can fluctuate from the ground-state Kramers doublet $\varepsilon_0$ into a conduction state near $E_{\rm F}$ and back to $\varepsilon_1$ or $\varepsilon_2$, involving singular quantum spin-flip transitions, as shown in Fig.~\ref{fig1}a. This process generates two narrow resonances~\cite{Schrieffer1966,Zamani2016}, that are shifted in energy by the final-state excitation energies, $\Delta_1$ and $\Delta_2$ where $\Delta_1 = \varepsilon_1 - \varepsilon_0 + \delta\Delta_1$ and $\Delta_2 = \varepsilon_2 - \varepsilon_0 + \delta\Delta_2$. Here, $\Delta_{1,2}^{(0)} = \varepsilon_{1,2} - \varepsilon_0$ are the bare CEF excitation energies, while $\delta\Delta_{1,2}$ are the additional many-body renormalizations~\cite{Reinert2001,Ehm2007,Pal2019}.

In this article, we report our observations on the CEF resonances of a Ce-based intermetallic germanide, CeAg$_{2}$Ge$_{2}$, hereafter referred to as CAG. The CAG compound undergoes a long-range antiferromagnetic ordering with the easy axis along [100] below 4.6\,K~\cite{Thamizavel2007,Thamizavel2008}. In the ground-state electronic configuration, the moments of the Ce ions create a spin-density-wave structure where the Ce spins fluctuate along the $b$-axis~\cite{Singh2011}. From high-resolution photoemission spectroscopy, the spin-orbit splitting between the 4$f_{7/2}$ and the 4$f_{5/2}$ is observed to be 280\,meV~\cite{Banik2014}, similar to other Ce-based compounds~\cite{Patthey1990}. In addition, CAG is reported to have a Kondo phase below a characteristic temperature of $\approx 3$\,K~\cite{Knopp1987,Banik2020}, which makes it a good example of Kondo-lattice antiferromagnet. While studies on the anisotropic physical properties of a single crystal of CAG have been reported before by means of electrical resistivity, magnetic susceptibility, specific heat, and magnetoresistance measurements~\cite{Thamizavel2007,Thamizavel2008}, there are no reports on the direct measurement of the underlying low-energy CEF states that play a succinct role in the hybridization. Inelastic neutron spectroscopy could only reveal the existence of a higher CEF state at 11\,meV~\cite{Knopp1987,Loidl1992}. In contrast, the heat-capacity measurement on CAG affirms the presence of low-energy CEF states~\cite{Thamizavel2008}. From the theoretical analysis of the magnetic susceptibility data, it has been predicted that the CEF states in CAG are separated by energy gaps of 5\,K and 130\,K~\cite{Thamizavel2007}, which clearly lie in the THz frequency range. We use THz time-domain spectroscopy to get direct and resonant access to the CEF states~\cite{Wetli2018,Pal2019,Yang2020,Yang2022,Yang2023}. We find two robust CEF states, namely CEF$_1$ and CEF$_2$ at 2.5\,meV (= 0.6\,THz) and 8.7\,meV (= 2.1\,THz), respectively, that show distinct temperature dependence. Our spectral weight analysis in the time domain further corroborates the corresponding temperature scales. In addition, we find that for temperatures less than 3\,K, the carrier scattering time diverges. We associate this divergence with the onset of strong electron-electron interactions that underpins the existence of the Kondo phase within the antiferromagnetic phase.

\section{Sample and Experimental method}
Single crystals of CeAg$_{2}$Ge$_{2}$ are grown by the self-flux method, where the binary eutectic composition of Ag-Ge is used as the flux. The starting materials, namely 99.9\% Ce, 99.999\% Ag, and 99.999\% Ge are mixed in the ratio of 1:16.25:6.75, placed in an alumina crucible, and then sealed in an evacuated quartz ampoule. The ampoule is then placed in a furnace at $1050^\circ$C for two days for complete blending of the solution. The furnace is then cooled down to the eutectic point of Ag-Ge flux ($650^\circ$C) over a period of three weeks, followed by rapid cooling to room temperature. The crystals are separated from the flux using a centrifuge. The compound crystallized in a body-centered tetragonal structure, similar to the ThCr$_{2}$Si$_{2}$ structure~\cite{Rauchschwable1985}, as shown schematically in Fig.~\ref{fig1}b. Samples from the same ingot have been used in an earlier report~\cite{Thamizavel2007} on electrical resistivity, magnetic susceptibility, magnetization, specific heat, and magnetoresistance.

The sample used in this work is cut from the ingot to orient the face perpendicular to the crystallographic $c$-axis. This surface is polished using a colloidal silica slurry, and it is ensured that the surface damages due to the polishing are within the submicron range (i.e., $\ll\lambda_{\rm THz}$, the wavelength of the THz radiation). The sample is then mounted in a temperature-controlled Janis SVT-400 helium reservoir cryostat. All our experiments are performed in a reflection geometry where linearly polarized THz radiation with a spectral range of $0.1-2.5$\,THz (= $0.4-10$\,meV) is used at an angle of incidence of $45^\circ$. The electric field of the incident THz radiation is oriented perpendicular to the crystallographic $a$-axis. 

Single-cycle THz pulses have been generated via optical rectification in a 0.5\,mm (110)-oriented ZnTe crystal, using 90\% of a Ti:Sapphire laser output (wavelength 800 nm, pulse duration 120\,fs, pulse repetition rate 1\,kHz, pulse energy 2\,mJ/pulse). The remaining 10\% of the fundamental pulse is used as a gating pulse for the free-space electro-optic sampling of the reflected THz radiation. The THz and the gating beams are collinearly focused onto a (110)-oriented ZnTe detection crystal. The THz-induced ellipticity of the gated light is measured using a quarter-wave plate, a Wollaston polarizer, and a balanced photodiode. The signal from the photodiode is then analyzed with a lock-in amplifier. The accessible time delay between the THz and the probe pulses can be significantly increased by suppressing the Fabry-P\'erot resonances from the faces of the 0.5\,mm (110)-oriented ZnTe crystal. This is achieved by extending the detection crystal with a 2\,mm THz-inactive (100)-oriented ZnTe crystal that is optically bonded to the back of the detection crystal. All measurements are performed in an inert nitrogen atmosphere to remove the water absorption lines. 

\section{Results and discussion}
\subsection{THz reflectance and conductivity} 
\begin{figure}[t!]
	\centering
	\includegraphics[width=\columnwidth]{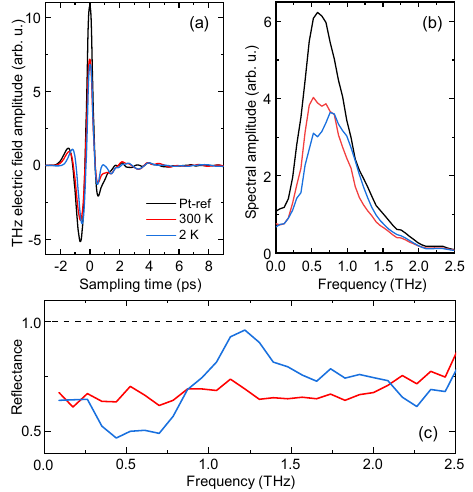}
	\caption{(a) THz electric field transients from the reference mirror and the CAG sample at two different temperatures. (b) Corresponding spectral amplitudes obtained via fast Fourier transformation of the time transients. (c) The magnitude of reflectance of the sample at 300\,K and 2\,K. The reflectance at 300\,K is flat and featureless. The minima in the 2\,K reflectance spectra on either side of the reflectivity peak correspond to the frequency regions of the CEF transitions. The dashed line represents 100\% reflectivity.}
	\label{fig2}
\end{figure}

The THz time transients reflected from the CAG sample are collected as a function of temperature. Figure~\ref{fig2}a shows such transients at two different temperatures, namely $T=300$\,K and $T=2$\,K. The reference in our experiments consists of a 15\,nm Pt film grown on a quartz substrate. This is placed next to the sample inside the cryostat so that sample $E_{\rm s}(t)$ and reference $E_{\rm r}(t)$ transients can be measured under identical experimental conditions. We note that the sample transient deviates from the reference transient where we see additional wiggles in the first few picoseconds, signaling the presence of resonance features. Figure~\ref{fig2}b shows the corresponding fast Fourier transforms of the time transients in Fig.~\ref{fig2}a. The reflectance spectra $r(\omega,T)$, shown in Fig.~\ref{fig2}c, are obtained from Fig.~\ref{fig2}b by taking a ratio of the sample spectrum at the respective temperatures $E_{\rm s}(\omega,T)$ to the room temperature reference spectrum $E_{\rm r}(\omega,300\,{\rm K})$, i.e.,
\begin{equation}\label{eq:1}
r(\omega,T) = \frac{E_{\rm s}(\omega,T)}{E_{\rm r}(\omega,300\,{\rm K})}.
\end{equation}
The spectrum at 300\,K, i.e., the red curve in Fig.~\ref{fig2}c displays a frequency-independent flat reflectance within the measured THz frequency range. Upon lowering the sample temperature to 2\,K, however, the spectrum develops two reflectance minima on either side of the reflectance maximum at around 1.2\,THz, see the blue curve in Fig.~\ref{fig2}c. We associate these minima to the many-body-renormalized CEF transitions $\varepsilon_0 \to \varepsilon_1$ and $\varepsilon_0 \to \varepsilon_2$, respectively. The first minimum, centered around 0.6\,THz, results in a conductivity peak that resembles a Lorentzian lineshape. The second minimum, in contrast, results in a conductivity peak at 2.1\,THz that, because of the interference with a co-existing phonon mode, creates a dip at 2.3\,THz. This case of Fano interference is addressed in the next section. The temperature-dependent DC conductivity in Fig. S1 of the supplementary information~\cite{supp}, shows that our sample is a good conductor. In addition, the zero-frequency extrapolation (discussed later) shows that the THz conductivity beautifully reconciles with the DC conductivity in the microwave region. This points to a low-frequency reflectance increase (known as the Hagen-Rubens response) in the microwave region. We could not access this range in our experiments and therefore observe the lower reflectance tail of the Hagen-Rubens response.

\begin{figure}[t!]
	\centering
	\includegraphics[width=\columnwidth]{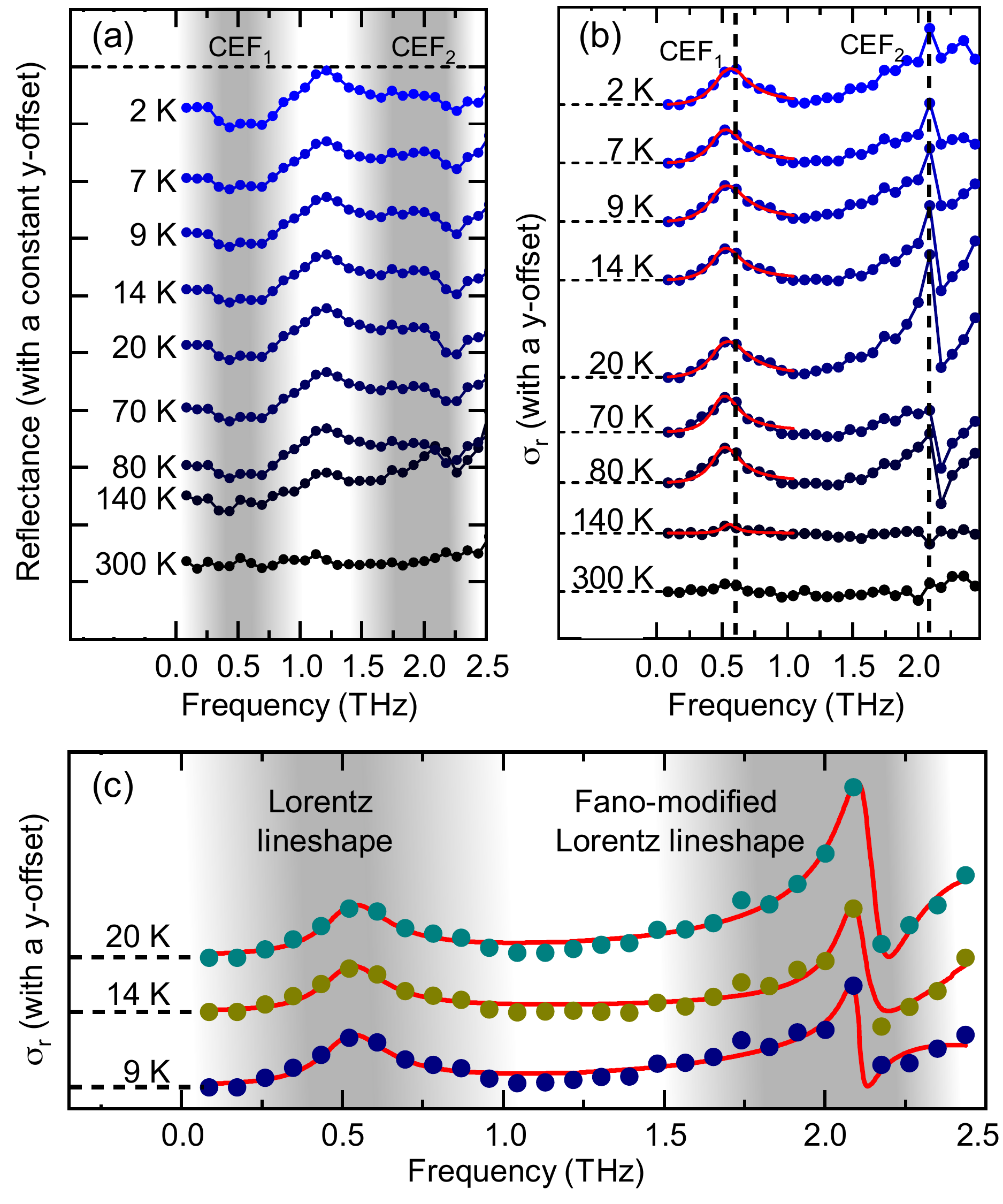}
	\caption{(a) Temperature-dependent THz reflectance. The gray-shaded regions show the CEF$_1$ and CEF$_2$ transitions, i.e., between $\varepsilon_0 \to \varepsilon_1$ and $\varepsilon_0 \to \varepsilon_2$, respectively. The horizontal dashed line represents 100\% reflectivity. All curves are displayed with a constant $y$-offset of 0.5 for each temperature values. (b) The frequency-resolved THz conductivity at different temperatures. The conductivity peaks at 0.6\,THz, and 2.1\,THz correspond to the two CEF transitions, as denoted by the vertical dashed lines. The horizontal dashed lines indicate the zero for each of the vertically shifted curves. The red lines are the theoretical curves obtained from the single-peak Lorentz oscillator model. (c) Complete modeling of the THz conductivity data using Eq.~(\ref{eq:5}) at three different temperatures. All conductivity values are renormalized with respect to the skin depth.}
	\label{fig3}
\end{figure}

To assign the CEF transitions, we thus evaluate the THz conductivity from the complex-valued reflectance in Eq.~(\ref{eq:1}). Using Fresnel's relations for an incident $p$-polarized configuration, the reflection coefficient $r(\omega)$ can be expressed as~\cite{Yang2020},
\begin{equation}\label{eq:2}
r(\omega) = \frac{n_2{\rm cos}(\theta_1) - n_1{\rm cos}(\theta_2)}{
n_2{\rm cos}(\theta_1) + n_1{\rm cos}(\theta_2)},
\end{equation}
where $n_1$ and $n_2$ are the refractive indices of air and sample, respectively. $\theta_1$ and $\theta_2$ are the angle of incidence and the angle of refraction, respectively. In our experimental setup, the incidence angle $\theta_1$ is $45^\circ$ and the refractive index of air is $n_1 = 1$. Using Snell's law ($n_1{\rm sin}\theta_1 = n_2{\rm sin}\theta_2)$, Eq.~(\ref{eq:2}) can be simplified as
\begin{equation}\label{eq:3}
r(\omega) = \frac{n_2^2(\omega) - \sqrt{2n_2^2(\omega)-1}}{n_2^2(\omega) + \sqrt{2n_2^2(\omega)-1}}.
\end{equation}
where $n_2(\omega)$ is the complex, frequency-dependent refractive index of the sample that is related to the complex dielectric constant as 
\begin{equation}\label{eq:4}
\epsilon(\omega)=\epsilon_{\rm r}(\omega)+i\epsilon_{\rm im}(\omega)=n_2^2(\omega).
\end{equation}
The optical conductivity being a linear response function, the real and imaginary parts of the THz optical conductivity can be obtained from the complex-valued dielectric constant using the relations~\cite{Basov2005,Basov2011}, $\sigma_{\rm r}(\omega) = \omega\epsilon_{0}\epsilon_{\rm im}(\omega)$ and $\sigma_{\rm im}(\omega) = -\omega\epsilon_{0}[\epsilon_{\rm r}(\omega)-1]$, where $\sigma(\omega)=\sigma_{\rm r}(\omega)+i\sigma_{\rm im}(\omega)$. The CEF transitions that give rise to spectral absorption in our sample get directly revealed in the real part of the THz conductivity ($\sigma_{\rm r}$), shown in Fig.~\ref{fig3}b for different temperatures. The corresponding temperature-dependent reflectance spectra are shown in Fig.~\ref{fig3}a. We assign the peaks observed at 0.6\,THz and 2.1\,THz to the CEF$_1$ and CEF$_2$ transitions, respectively, that is in close agreement with the predicted values~\cite{Thamizavel2007} and scrutinize these features further in the next sections. Note that both CEF$_1$ and CEF$_2$ appear below $T\le140$\,K. These resonance features evidently display a robust temperature dependence as shown in Fig.~\ref{fig3}b. Both the real part and the imaginary part (shown in the Fig.~S3b of the supplementary material~\cite{supp}) of the THz conductivities reconcile with the DC conductivity upon renormalization with respect to the skin-depth, which is in the order of a micron. This is achieved by performing the zero-frequency extrapolation of the THz conductivity~\cite{Bosse2016}, see Fig.~S3 in the supplementary material~\cite{supp}. We find that the Drude tail is significantly modified by the presence of low-frequency modes and the Drude peak is pushed more into the microwave regime that, as mentioned, goes beyond our experimental frequency range. The imaginary THz conductivity also shows an excellent agreement with the extrapolated Drude tail.

\begin{figure}[t!]
	\centering
	\includegraphics[width=\columnwidth]{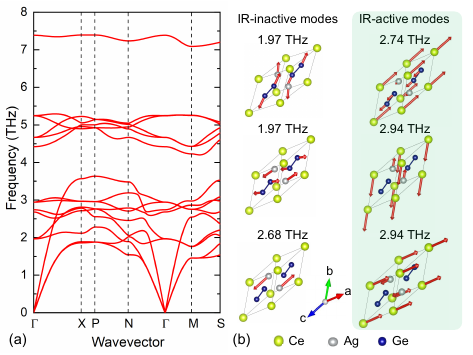}
	\caption{(a) The calculated phonon dispersion  for CeAg$_{2}$Ge$_{2}$. (b) Relevant infrared (IR)-inactive and IR-active phonon modes at the $\Gamma$-point. The 1.97\,THz IR-inactive and the 2.94\,THz IR-active modes are degenerate. The 2.74\,THz IR-active mode is in close proximity to the CEF$_2$ transition.}
	\label{fig4}
\end{figure}

\subsection{Modelling of THz conductivity}
The two CEF transitions show very distinct and different spectral lineshapes. It can be seen from Fig.~\ref{fig3}b that the CEF$_1$ transition represents a typical Lorentzian lineshape that can be explained using a single-peak Lorentz oscillator model. In contrast, the presence of a broad reflectance response from $1.2-2$\,THz and a minimum at 2.3\,THz (see Fig.~\ref{fig3}a) significantly modifies the THz conductivity spectrum of CEF$_2$ transition to reveal a \emph{Fano-modified Lorentz} lineshape. From the phonon dispersion obtained using first-principles calculation as shown in Fig.~\ref{fig4}a, we find two infrared (IR)-active phonon mode at the $\Gamma$-point (zone-centre) below 3\,THz: the $A_{2u}$ mode at 2.74\,THz and the $E_{u}$ mode 2.94\,THz (Refer to Appendix~A  for further details on the phonon calculations), which the THz radiation can suitably probe. Both these phonon modes, shown in Fig.~\ref{fig4}b, are in close proximity to the CEF$_2$ transition. This leads to quantum interference between the correlated electrons and phonons, manifesting as a Fano-resonance at 2.1\,THz. Similar Fano-physics have been reported before in topological materials~\cite{LaForge2010,Sim2015}, Weyl semimetals like TaAs and NbAs~\cite{Xu2017,Coulter2019} and in strongly correlated electronic systems like stripe-phase nikelates~\cite{Coslovich2013}, and high $T_c$ superconductors~\cite{Xu2015,Homes2016}, where considerable contributions from coupling between phonons and conventional fermions have been observed. This feature, however, weakens as we lower the temperature so that the CEF resonance at 2.1\,THz comes to the fore. In contrast, there are no IR-active phonon modes below 1\,THz or in close proximity to the CEF$_1$ transition, as a result of which the CEF$_1$ transition at 0.6\,THz remains unaffected. The THz conductivity can thus be phenomenologically modeled using a combination of single-peak Lorentz oscillator function~\cite{Lloyd-Hughes2012} and a Fano-modified Lorentz function~\cite{Chen2014}, given by
\begin{equation}\label{eq:5}
\begin{split}
    \sigma_{\rm L}^{\rm r}(\omega) &= \sigma_{\rm L}^{\rm r,CEF_1}(\omega) + \sigma_{\rm L}^{\rm r,CEF_2}(\omega)\\
    &= \frac{n_ee^2\omega^2}{m^*}\frac{\tau_1}{\omega^2+\tau_1^2(\omega_1^2-\omega^2)^2}\\
    &+ \frac{C[(\omega-\omega_2)\tau_2-2\pi q]^2}{4\pi^2+(\omega-\omega_2)^2\tau_2^2}\cdot\frac{(2\pi\gamma)^2}{(2\pi\gamma)^2+(\omega-\delta)^2},
\end{split}
\end{equation}
where $n_e$, $e$, and $m^*$ are the density, charge, and the effective mass of the electrons, respectively. $C$ is a temperature-dependent constant that accounts for the corresponding spectral weight and $q$ is the asymmetry parameter. $\omega_1$ and $\omega_2$ are the peak frequencies while $\tau_1$ and $\tau_2$ are the scattering times of the respective CEF transitions. $\delta$ represents the minimum of the shifted reflectance dip while $\gamma$ is the corresponding width of the dip. Figure~\ref{fig3}c shows the modeled curves (as per Eq.~(\ref{eq:5})) in red that clearly display a beautiful agreement with the experimental data at three exemplary temperatures. Since the CEF$_2$ transition is modified due to the background electronic scatterings, we turn our focus to the CEF$_1$ transition that purely resembles the interplay of the hybridization between the 4$f$-electrons and the conduction electrons.

\begin{figure}[t!]
	\centering
	\includegraphics[width=\columnwidth]{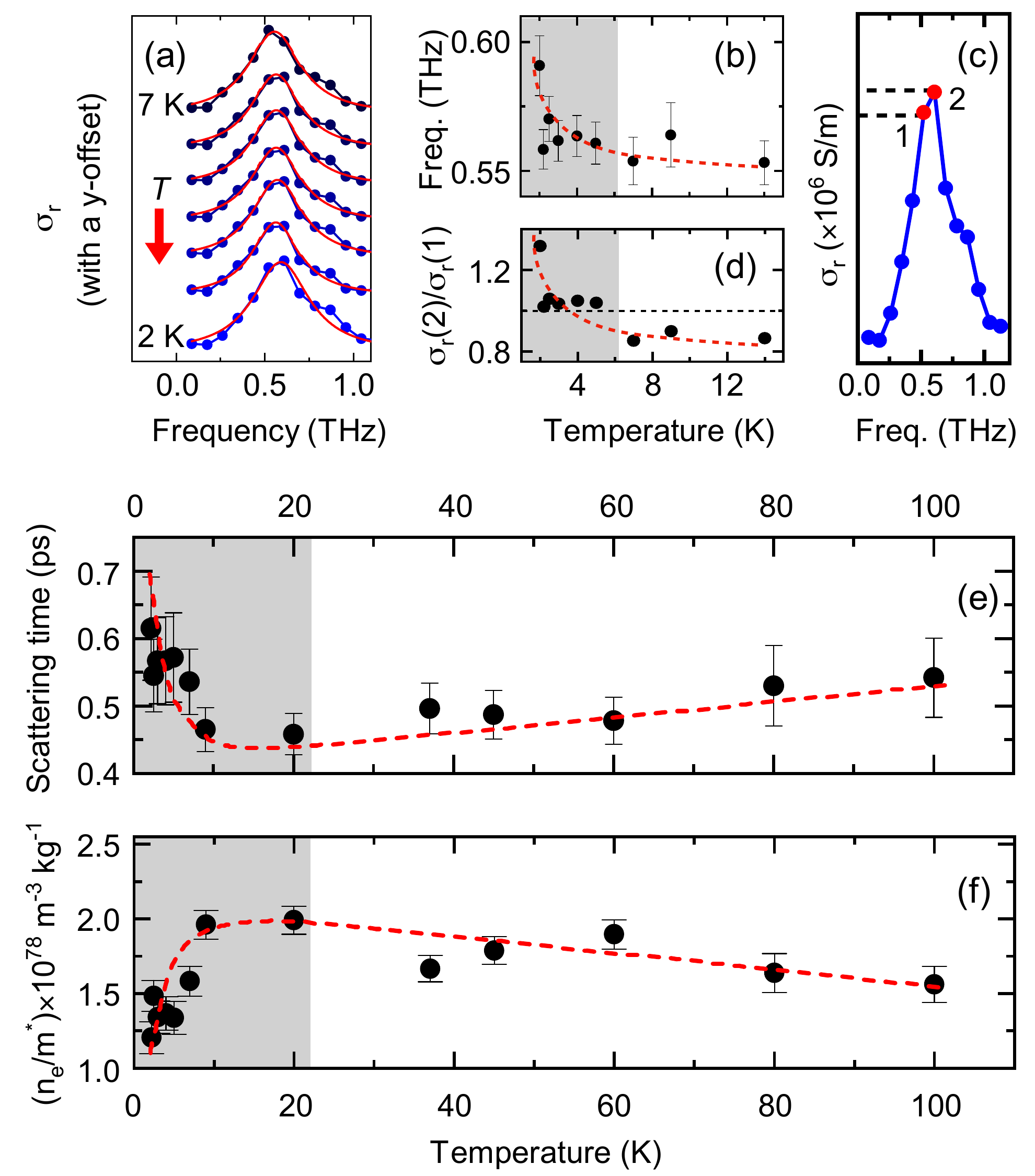}
	\caption{Parameters from the theoretical Lorentz modeling. (a) The real part of the THz conductivity corresponding to the CEF$_1$ transition. The red curves represent the fitted single-peak Lorentz function. (b) The CEF$_1$ peak frequency extracted from (a). There is an evident blue-shift of 0.1\,THz of the peak frequency as the sample temperature is lowered below 5\,K, i.e., when the sample enters the antiferromagnetic phase (indicated by the gray region and the red-dashed guide-to-the-eye line). (c) Conductivity spectrum at 2\,K, where the points marked as 1 and 2 correspond to the conductivity value at 0.52\,THz and 0.61\,THz, respectively. The horizontal-dashed line indicates $\sigma_r(2)=\sigma_r(1)$, and hence the crossover point. (d) The blue-shift calculated by taking a ratio of the points marked as point 2 ($\sigma_r(2)$) and 1 ($\sigma_r(1)$), and indicated using the red-dashed guide-to-the-eye line. (e) The temperature dependence of the scattering time with respect to the CEF$_1$ transition. (f) The temperature dependence of $n_e/m^*$, corresponding to the first CEF transition. The gray-shaded regions in (e) and (f) correspond to the low-temperature divergence region. The error bars in all the plots indicate the standard errors of the respective parameters obtained by fitting the first term of Eq.~(\ref{eq:5}) with the data. These error bars are not statistical errors, but overall systematic errors introduced by the parameter fits while modelling the conductivity data. The red-dashed guide-to-the-eye lines show the temperature-dependent behavior of the scattering time and $n_e/m^*$.}
	\label{fig5}
\end{figure}

\subsection{The CEF$_1$ transition}
The red curves in Fig.~\ref{fig3}b represent the modeled value of $\sigma_{\rm L}^{\rm r,CEF_1}(\omega)$ at different temperatures that to a very good extent reproduces the experimentally-obtained THz conductivity below 140\,K. The parameters, such as the center frequencies of the CEF transitions, the corresponding scattering times, and the ratio of electron density to the effective electronic mass, extracted from the modeled data along with their temperature dependence, are shown in Fig.~\ref{fig5}. The temperature dependence of the center frequency corresponding to CEF$_1$ displays a robust behavior until 7\,K, below which the peak shifts to a higher frequency until the lowest temperature measured. Figure~\ref{fig5}a shows the real part of complex conductivity along with the fitted single-peak Lorentzian function used to extract the CEF$_1$ peak frequency. The temperature dependence of the extracted peak frequency is plotted in Fig.~\ref{fig5}b. As the temperature is lowered from 7\,K, a blue-shift of $\approx 0.1$\,THz is recorded until the lowest temperature measured. To scrutinize the temperature dependence of the peak frequency more closely, we have also taken a ratio of $\sigma_r$ at the two corresponding peak maxima (see Figs.~\ref{fig5}c and~\ref{fig5}d). More precisely, the CEF peak blue-shifts once the sample enters into the antiferromagnetic phase, i.e., $T\leq5$\,K where $T_{\rm N}=4.6$\,K, see Fig.~\ref{fig5}b. The low-temperature magnetic susceptibility measurements~\cite{Thamizavel2007}, previously performed on CAG, revealed that the sample has a strong magnetic anisotropy. From the theoretical CEF analysis of the susceptibility data, one already finds a strong correlation between the CEF parameters and the magnetic anisotropy. Our experiments present beautiful experimental evidence of how the low-energy CEF states are affected as the system enters into a magnetically-ordered state. Specifically, owing to the competitive energy scales of the Kondo effect and the RKKY interaction, the onset of the magnetic ordering renormalizes the CEF states. We speculate that the RKKY-assisted renormalization~\cite{Nejati2017} leads to an increase of $\Delta_1$ and ultimately results in the blue-shift of the CEF$_1$ transition. In other words, the RKKY-assisted renormalization additionally shifts the final-state excitation energy by $\Delta\Omega$, such that $\Delta_1 = \Delta_1^{(0)}+\delta\Delta_1+\Delta\Omega$, where $\Delta_1^{(0)}=\varepsilon_1-\varepsilon_0$. Evidence of such blue-shifts has been previously reported in $\text{CeCu}_5\text{Au}$ which has an antiferromagnetic ground state~\cite{Yang2020}.

\begin{figure}[t!]
    \centering
	\includegraphics[width=\columnwidth]{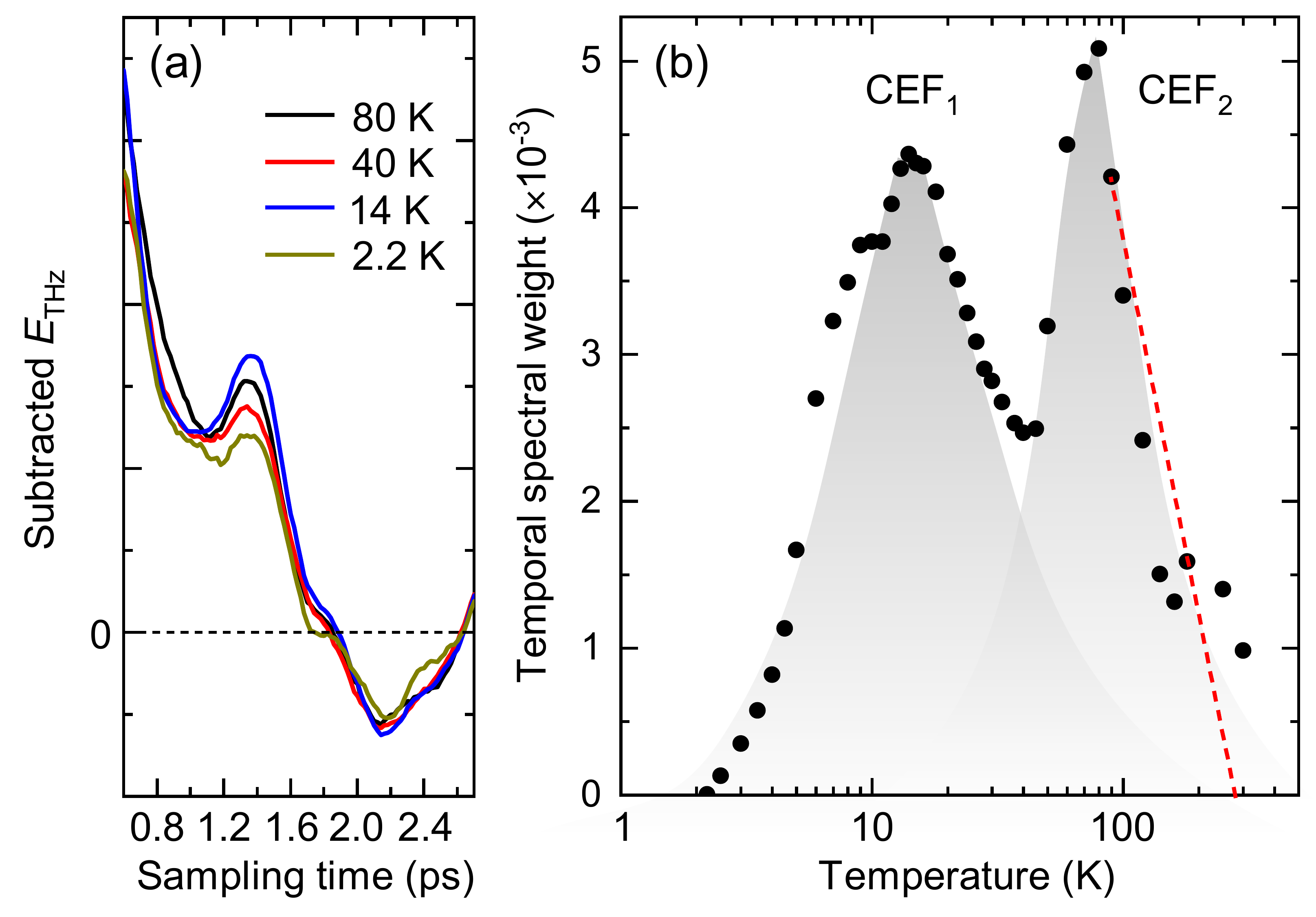}
	\caption{(a) Subtracted THz time traces at several temperatures. (b) The temporal spectral weights obtained by integrating the corresponding to the exemplary time traces in (a) within the time window [0.6\,ps, 2.7\,ps].  The red-dashed line indicates the high-temperature logarithmic increase. The absolute values of the data experience shift up to 20\%, subject to the experimental conditions, i.e., the number of scans and a small shift of the integration window. This however does not affect the extracted temperature scales corresponding to the CEF transitions.}
	\label{fig6}
\end{figure}

Figure~\ref{fig5}e shows the bound-carrier scattering times corresponding to the first CEF transition. The scattering times are in the sub-picosecond scales with an appreciable temperature dependence. It can be followed from the guide-to-the-eye line that the scattering time ($\tau_1$) decreases  upon decreasing the temperature to 20\,K. This is a characteristic feature indicating the presence of electron-phonon coupling, where the scattering times associated with electron-phonon interactions and electron-electron interactions are both on the order of a few picoseconds~\cite{Groeneveld1995,Degiorgi1999,Demsar2003}. Upon decreasing the temperature further, we see low-temperature divergence in the scattering times. This results from the fact that the scattering times associated with the electron-electron interactions start to dominate over the electron-phonon interactions. This is a clear indication of the Kondo-like hybridization of the $4f$-electrons with the conduction electrons. The electron density to effective mass ratio ($n_e/m^*$), further, displays a behavior complementary to the scattering time (see Fig.~\ref{fig5}f), where the ratio increases as we decrease the sample temperature up to 10\,K, followed by a decrease in the ratio. The decrease in the ratio either implies an increase in effective mass or a decrease in the carrier density, both signaling the onset of hybridization~\cite{Degiorgi1999}. Alternatively, we also extracted the width and the integral of the CEF$_1$ absorption peaks at several temperatures and plotted them in Fig.~S4 of the supplementary material~\cite{supp}, where we show the reconciliation of the temperature behavior of both the scattering time and the $n_e/m^*$ ratio.

\subsection{THz temporal weight analysis}
The relevant energy/temperature scales associated with the Kondo effect and the CEF transitions reveal themselves as temporally separated distinct features within the time transients and the corresponding temporal weights~\cite{Wetli2018,Pal2019}. Here, the temporal weight is a measure of the momentum-integrated spectral density corresponding to the CEF states. To this end, we carried out the temporal-weight analysis of the temperature-dependent time transients. From our discussion above, the two CEF transitions are located at 0.6\,THz (2.5\,meV) and 2.1\,THz (8.7\,meV) that in the temperature scale would correspond to 24\,K and 100\,K, respectively. Using the scale~\cite{Wetli2018,Pal2019} $\tau_{\rm CEF} = h/k_{\rm B}T_{\rm CEF}$, we expect the two CEF transitions to be temporally located in the tail of the instantaneous response, i.e., in the range [0.5\,ps, 2.0\,ps]. 

Following a similar procedure as highlighted in Ref.~\cite{Pal2019}, we first subtract the high-temperature data from all traces to remove any incoherent contributions to the spectral weight. Next, we average all the low-temperature ($T\leq4$\,K) time traces and subtract the resulting curve from all the traces. At low temperatures, the occupation of CEF states freezes leaving behind only the stimulated single-particle response of the conduction electrons, which appears in the instantaneous response. The subtraction of low-temperature time traces removes these additional contributions leaving behind the pure response from the CEF states in the resultant traces. Figure~\ref{fig6}a shows these resultant time traces at a few selected temperatures. We then integrate the time traces of the squared THz electric field over the interval [0.6\,ps, 2.7\,ps]. The temperature-dependent temporal weights thus obtained are shown in Fig.~\ref{fig6}b. The THz excitation being sensitive to the occupied states only, the weights calculated in this manner represent directly the total occupation numbers of the CEF states near $E_{\rm F}$. The choice of the integration window is such that the temporal response from both CEF states is considered. Further, the lower boundary is chosen to capture the maximum possible range of the CEF signal while excluding the tail of the instantaneous pulse~\cite{Yang2022}, which exists significantly up to 0.5\,ps. This limits the lower boundary of the window to be 0.6\,ps instead of 0.5\,ps.

The spectral weight in Fig.~\ref{fig6}b shows a very distinct temperature dependence. Upon cooling the sample from room temperature, we first see a logarithmic increase (the red-dashed line in Fig.~\ref{fig6}b) of the spectral weight that signals the high-temperature Kondo scale stemming from the excited CEF states~\cite{Pal2019}. The peak observed in Fig.~\ref{fig6}b at around 100\,K corresponds to the occupation of the second CEF state. This is followed by a decrease of the signal and then a second peak at around 15\,K, corresponding to the occupation of the first CEF state. Both temperature scales of the spectral weight in the time domain beautifully corroborate with the energy scales of the CEF states obtained in the frequency domain. At very low temperatures, the overall spectral weight from the temporal CEF features vanishes, in agreement with previous observations in heavy-fermion CeCu$_{6-x}$Au$_{x}$ samples~\cite{Pal2019}.

\section{Conclusion}
To summarize, we have identified the low-energy CEF transitions in single crystals of CeAg$_{2}$Ge$_{2}$, which is a typical example of a Kondo-lattice antiferromagnet. We used time-domain THz reflection spectroscopy to evaluate the THz conductivity and characterize the temperature dependence of the CEF transitions. We found peaks at 0.6\,THz and 2.1\,THz in the conductivity spectra that correspond to the two low-energy CEF transitions. In addition, our results highlight the competition between the Kondo and the RKKY interactions, leading to additional renormalization in the energies of the CEF states, manifesting itself as a blue shift of the CEF$_1$ transition in the antiferromagnetic phase. We also found that the scattering times corresponding to the CEF$_1$ transition show a low-temperature divergence that is characteristic of the electron-electron correlations underpinning the Kondo-like hybridization of the Ce-$4f$ electrons and the conduction electrons. At higher temperatures, however, the scattering time varies linear-in-temperature, indicating the presence of electron-phonon coupling. Note that the values obtained from our direct measurements of the CEF transitions are slightly different compared to previously reported ones~\cite{Thamizavel2007}. This is, however, not surprising because the reported values are not directly measured but obtained indirectly from the CEF analysis of the magnetization data. We also report the presence of low-energy IR-active phonon modes in this material and the coupling to the CEF$_2$ transition that manifests as a Fano-modified Lorentz lineshape of the latter. In the temporal spectral weight analysis, we were able to corroborate the temperature scales for the CEF transitions. Our experiments provide the first direct evidence of the underlying low-energy CEF states present in a Kondo-lattice antiferromagnet. The understanding gained from our experiments apparently opens up the possibility to drive the CEF transitions in a nonlinear fashion using multiple THz pulses and gain a deeper understanding of the underlying hybridization at ultrafast timescales~\cite{Yang2023}.

\section*{Acknowledgement}
P.S. and S.P. acknowledge the support from DAE through the project Basic Research in Physical and Multidisciplinary Sciences via RIN4001. S.P. also acknowledges the start-up support from DAE through NISER and SERB through SERB-SRG via Project No.~SRG/2022/000290. C.-J.Y. and M.F. acknowledge the financial support from NCCR MUST via No.~PSP~1-003448-051. M.F. acknowledges the support from the SNSF through Project No.~200021\_178825. A.K.N. acknowledges the support from SERB through SERB-SRG via Project No.~SRG/2019/000867.\\

\begin{table}[b!]
 \centering
 \caption{Irreducible representations (Irreps) of phonon modes at the $\Gamma$-point and their IR activity is listed.\\}
\begin{tabular}{ccccc}
\hline \hline
Frequency & & Irreps & & IR-active\\
(THz) & &  & & \\
\hline
  &  &  &  &   \\
1.97  &  & $E_g$ & & No  \\ 
2.68  &  & $B_{1g}$& & No     \\
2.74  &  & $A_{2u}$& & Yes     \\  
2.94  &  & $E_u$& & Yes   \\
4.42  &  & $A_{2u}$& & Yes    \\
4.67  &  & $E_u$ & & Yes    \\
5.25  &  & $E_g$ & & No    \\
7.39  &  & $A_{1g}$& & No    \\
\hline \hline
\end{tabular}
\label{tab:ir}                                  
\end{table}

\section*{Appendix A: First-principle calculations}
Density-functional theory calculations are performed within the Quantum Espresso package~\cite{Giannozzi2017jpcm,Giannozzi2009jpcm} where projector-augmented wave method~\cite{Kresse1999prb,Bloechl1994prb} and a plane-wave basis set are employed. The generalized-gradient approximation with Perdew-Burke-Ernzerhof functional form is used for the calculation of exchange-correlation energy~\cite{Perdew1996prl}. Pseudopotentials with nonlinear core corrections are considered with semi-cores $s$, $p$, $d$, while semi-core $d$ state is treated as valence states for Ce and Ge atoms, respectively. The kinetic energy cutoff for plane-wave functions of 60\,Ry and charge density of 720\,Ry are considered in our calculations. We start with the tetragonal crystal structure (space group: $I4/mmm$, No. 139) of CeAg$_2$Ge$_2$. The conventional 10-atom unit cell with $|{\bf a}| = |{\bf b}| = 4.303$\,\AA{}, $|{\bf c}| = 10.974$\,\AA{} was fully optimized with energy and force convergence criterion of 4 $\times$ 10$^{-9}$\,Ry and 7 $\times$ 10$^{-5}$\,Ry/Bohr, respectively followed by the phonon calculations. A 10$\times$10$\times$4 uniform $k$-mesh and Methfessel-Paxton smearing are considered in our calculations. The optimization leads to small changes ($<$ 1.1 \%) in the lattice constants. The phonon dispersion is calculated for a primitive unit cell of 5 atoms with the finite differences approach considering 2 $\times$ 2 $\times$ 2 supercell. The post-processing of phonon calculations is done using Phonopy~\cite{Togo2023jpcm,Togo2023jpsj} software. The Infrared (IR) active modes are analyzed using the SAM module~\cite{Kroumova2003pt} of the Bilbao Crystallographic Server. 

The calculated phonon dispersion plot for the primitive cell of CeAg$_2$Ge$_2$ is shown in Fig.~\ref{fig4}(a). The mode analyses~\cite{Kroumova2003pt} at the $\Gamma$-point using irreducible representations is presented in Table~\ref{tab:ir}. It is evident that the infrared-active (IR-active) modes related to Ce, Ag, and Ge atoms (situated at the 2$a$, 4$d$, and 4$e$ Wyckoff positions, respectively) possess two allowed representations, namely $A_{2u}$ and $E_u$. One can clearly see that below 3\,THz there are only two IR-active phonon modes: the $E_u$ mode at 2.94\,THz and the $A_{2u}$ mode at 2.74\,THz, which are in close proximity to the CEF$_2$ transition. All other modes below 3\,THz are IR-inactive and not accessible by the THz radiation.

\end{document}